\documentclass{article}
\usepackage{spconf,amsmath,amssymb,graphicx,hyperref}
\hypersetup{hidelinks}
\usepackage[capitalize]{cleveref}
\usepackage{booktabs}
\usepackage{multirow}
\usepackage{flafter}   
\usepackage{xcolor}
\usepackage{pgfplots}
\pgfplotsset{compat=newest}
\definecolor{cNSF}{HTML}{0072B2}
\definecolor{cAMI}{HTML}{E69F00}
\definecolor{cMx6}{HTML}{009E73}
\definecolor{cSBC}{HTML}{CC79A7}

\newcommand{\dicow}{DiCoW}          
\newcommand{\dicop}{DiCoP}          
\newcommand{\cpwer}{cpWER}
\newcommand{\tsasr}{TS-ASR}
\newcommand{\mtasr}{MT-ASR}
\newcommand{\csa}{CSA}              

\newcommand{\rev}[1]{{#1}}     
\newcommand{\tcite}[1]{\textsuperscript{\scriptsize\cite{#1}}}  

\title{Pandemonium: Longform Target-Speaker Speech Recognition with Cross-Speaker Interactions}

\name{\begin{tabular}{@{}c@{}}
  Alexander Polok$^{1}$, Dominik Klement$^{1}$, Ivan Medennikov$^{2}$, Antoine Laurent$^{3}$, \\
  Petr P\'{a}lka$^{1}$, Samuele Cornell$^{4}$, Shinji Watanabe$^{4}$, Sanjeev Khudanpur$^{5}$, \\
  Jagadeesh Balam$^{2}$, Jan \v{C}ernock\'{y}$^{1}$, Luk\'{a}\v{s} Burget$^{1}$, Matthew Wiesner$^{5}$
  \end{tabular}%
  \thanks{
  This work was supported by JSALT 2026 at Johns Hopkins University with funds from NSF CCRI Grant No. 2120435, Google DeepMind, JHU HLTCOE, JHU AI2AI, and ACL. Additional support was provided by the Ministry of Education, Youth and Sports of the Czech Republic (MoE) through the OP JAK project ``Linguistics, Artificial Intelligence and Language and Speech Technologies: from Research to Applications'' (ID: CZ.02.01.01/\allowbreak 00/\allowbreak 23\_020/\allowbreak 0008518) and the Brno Ph.D. Talent Scholarship Programme. Computational resources were provided by the e-INFRA CZ project (ID: 90254) and IDRIS under GENCI allocations AD011012527, A0201012177, and AD011016176. Generative AI tools were used to help revise and refine the manuscript.}
}

\address{
  $^{1}$Brno University of Technology, Czech Republic \quad
  $^{2}$NVIDIA, USA \quad
  $^{3}$pyannoteAI, France \\
  $^{4}$Carnegie Mellon University, USA \quad
  $^{5}$Johns Hopkins University, USA}

\begin{document}
\ninept
\setlength{\abovedisplayskip}{4pt}
\setlength{\belowdisplayskip}{4pt}
\setlength{\abovedisplayshortskip}{2pt}
\setlength{\belowdisplayshortskip}{2pt}
\maketitle

\begin{abstract}
Diarization-conditioned target-speaker ASR (\tsasr{}), in particular Diarization-conditioned Whisper (\dicow{}), is an accurate and data-efficient approach to multi-talker transcription.
It has, however, three structural limitations: the model transcribes each speaker \emph{independently}, operates on short (30\,s) windows, and optimizes its diarization and \tsasr{} components separately.
We address the first two.
\rev{We integrate the diarization conditioning into a longform Parakeet backbone (\dicop{}) and add \emph{cross-speaker attention}, which lets the model attend across the speaker dimension so that speakers are no longer decoded in isolation, and an \emph{STNO mask refiner} that re-estimates the diarization-derived conditioning.}
\rev{As conversational data with genuine multi-speaker interaction is scarce, we scale training with a pseudo-labeling pipeline that mines 5.4k hours of in-the-wild multi-speaker audio from YODAS2.}
Combining these with synthetic and real data yields state-of-the-art results across several meeting benchmarks.
We open-source the resulting models---\dicop{} and \emph{Pandemonium}---and the pseudo-labeled data.
\end{abstract}


\begin{figure}[t]
\centering
\includegraphics[width=0.65\columnwidth]{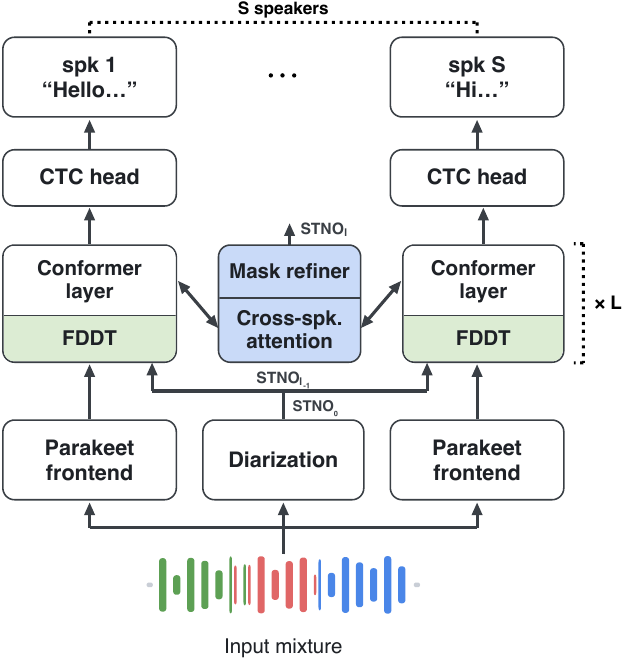}
\vspace{-7pt}
\caption{Overview of \emph{Pandemonium}.
A diarizer yields per-speaker masks that condition (FDDT) a longform Parakeet encoder processing the $S$ speakers as a batch.
\rev{Cross-speaker attention attends across speakers and a mask refiner sharpens the ambiguous STNO conditioning (blue boxes); a shared CTC head decodes each.}}
\label{fig:system}
\end{figure}

\section{Introduction}
\label{sec:intro}
Transcribing natural conversations remains challenging because of speaker overlap, diverse acoustic conditions, colloquial language, an unknown and varying number of participants, and recordings that span minutes rather than seconds.
\rev{Multi-talker ASR (\mtasr{}) transcribes all speakers with a single model.}
\rev{One line of work, serialized output training (SOT), serializes all speakers into a single output stream; it was introduced with attention-based encoder-decoders~\cite{kanda20b_interspeech} and has recently been adopted by LLM-based systems~\cite{shi2025serialized,peng2026vibevoiceasrtechnicalreport}.}
\rev{Others emit several streams under a permutation-invariant objective~\cite{yu17b_interspeech} or build on source separation~\cite{niu24_chime}.}
\rev{In contrast, target-speaker ASR (\tsasr{}) conditions a single-speaker recognizer on one speaker at a time and emits only that speaker's transcript.}
\rev{It is run once per detected speaker to reconstruct the full speaker-attributed transcript~\cite{ts_asr_kanda,speaker_beam}.}
Across backbones and data budgets, \tsasr{} has proven accurate and data-efficient~\cite{park2026pushingboundariesstreamingmultispeaker}.
Our prior work, Diarization-conditioned Whisper (\dicow{})~\cite{dicow}, is representative of \tsasr{}: it conditions the model on per-speaker, per-frame activities derived from a diarizer.
It has transferred to E-Branchformer~\cite{ebranchformer}, Whisper~\cite{radford2023robust,dicow}, and a spoken LLM~\cite{dixtral}, so it is largely backbone-agnostic.

\rev{Despite these strengths, \dicow{} has three structural limitations.}
\textbf{(i)} It models each speaker \emph{independently}: although the other speakers are present in the input audio, the model has no access to how they were encoded or what was decoded for them, discarding the cross-speaker cues that help resolve overlaps, turn-taking, and back-channels.
\textbf{(ii)} It is trained and decoded on short, fixed \emph{30\,s windows} inherited from the Whisper backbone, which caps the context available at both training and inference time.
\textbf{(iii)} It runs a \emph{multi-stage pipeline} in which diarization and recognition are trained separately.
In this work we address the first two limitations.

\rev{Recent longform ASR backbones~\cite{shi2026qwen3asrtechnicalreport}---in particular the Parakeet family of FastConformer models~\cite{sekoyan2025canary1bv2parakeettdt06bv3efficient,fast_conformer}---combine aggressive convolutional subsampling with local window attention and global tokens, processing whole meetings~\cite{vinnikov24_interspeech, Mccowan2005_ami, Yu2021M2MetTI, fu21b_aishell} on a single GPU without chunking.}
\rev{Meanwhile, systems that decode all speakers jointly (SOT~\cite{kanda20b_interspeech} and its LLM-based successors~\cite{shi2025serialized, peng2026vibevoiceasrtechnicalreport}) show that joint access to all speakers improves transcripts, yet they generalize poorly to unseen speaker counts and heavy overlap.}

\rev{We integrate the diarization conditioning of \dicow{} into Parakeet (\dicop{}), using its local attention and a CTC loss to train faster and decode whole conversations in one pass at a very low real-time factor (\cref{subsec:longform,subsec:ctc}), and design a new interaction module for \tsasr{}: \emph{cross-speaker attention} (\csa{}), which lets the model attend across the speaker dimension, complemented by a mask refiner that sharpens the diarization-derived conditioning (\cref{sec:csa}).}
\rev{Since such interaction is scarce in curated corpora, we mine multi-speaker recordings from YODAS2~\cite{yodas} and pseudo-label them with \dicow{}, obtaining 5.4k hours of conversational data (\cref{sec:pseudo}).}
Training on synthetic, pseudo-labeled, and real data gives our final system, \emph{Pandemonium} (\cref{fig:system}).\footnote{Pandemonium: the collective noun for a flock of parakeets, or chaos.}

\noindent\textbf{Contributions.}
We release:\footnote{\url{https://hf.co/BUT-FIT}}
\textbf{(i)} \dicop{}, a longform diarization-conditioned Parakeet \tsasr{} in TDT and faster CTC variants;
\textbf{(ii)} 5.4k\,h of pseudo-labeled multi-speaker YODAS2 conversations (\cref{sec:pseudo});
\textbf{(iii)} \emph{Pandemonium}, \dicop{} with \csa{} and an STNO mask refiner, with state-of-the-art results (\cref{tab:main}).

\begin{table}[!t]
\centering
\caption{Main results, \cpwer{}~[\%]~$\downarrow$.
\dicop{}: longform Parakeet backbone (loc/TDT of \cref{fig:tradeoff}); \emph{Pandemonium} adds \csa{} and the STNO refiner, trained synthetic\,$\to$\,pseudo\,$\to$\,real.
\rev{+\,MOVER: all distant microphones decoded separately, merged with MOVER~\cite{kamo25_interspeech} (SBC: stereo, unchanged).}
\rev{SOTA: best published single-microphone result (Sep 2026).}
Best single-microphone result per block in \textbf{bold}; abbreviations in \cref{subsec:datasets}.
}
\label{tab:main}
\setlength{\tabcolsep}{2pt}
\footnotesize
\begin{tabular}{@{}l cc cccc c@{}}
\toprule
 & \multicolumn{2}{c}{In-domain} & \multicolumn{4}{c}{Out-of-domain} & \\
\cmidrule(lr){2-3}\cmidrule(lr){4-7}
System & NSF & AMI & ICSI & Mx6 & SBC & DiPCo & Avg \\
\midrule
\multicolumn{8}{@{}l}{\emph{Oracle diarization}} \\
\dicow{}~v3.3~\cite{polok2026sedicow} & 15.5 & 14.5 & 14.1 & 11.4 & 15.0 & 27.1 & 16.3 \\
SE-\dicow{}~\cite{polok2026sedicow}   & 15.3 & 14.4 & 14.3 & 11.7 & 15.2 & 26.2 & 16.2 \\
\dicop{} (no \csa{})           & 17.3 & 14.5 & 14.4 & 11.9 & 16.0 & 26.1 & 16.7 \\
\textbf{Pandemonium}           & \textbf{14.1} & \textbf{13.4} & \textbf{13.3} & \textbf{11.1} & \textbf{13.8} & \textbf{22.4} & \textbf{14.7} \\
\rev{\;\;+\,MOVER} & \rev{10.8} & \rev{12.6} & \rev{11.9} & \rev{9.7} & \rev{13.8} & \rev{19.4} & \rev{13.0} \\
\midrule
\multicolumn{8}{@{}l}{\emph{Predicted diarization (pyannoteAI Precision-3)~\cite{laurent26_odyssey,bredin23_interspeech}}} \\
\dicow{}~v3.3~\cite{polok2026sedicow} & 23.0 & 17.3 & 18.6 & \textbf{11.5} & 19.3 & 30.0 & 20.0 \\
SE-\dicow{}~\cite{polok2026sedicow}   & 22.8 & 17.3 & 18.2 & 11.8 & 19.6 & 29.3 & 19.8 \\
\dicop{} (no \csa{})           & 23.2 & 16.9 & \textbf{18.0} & 12.1 & 19.4 & 28.9 & 19.8 \\
\textbf{Pandemonium}           & \textbf{22.4} & \textbf{16.8} & 18.1 & 11.7 & \textbf{18.6} & \textbf{27.4} & \textbf{19.2} \\
\rev{\;\;+\,MOVER} & \rev{19.0} & \rev{14.5} & \rev{15.7} & \rev{10.2} & \rev{18.6} & \rev{24.8} & \rev{17.1} \\
\midrule
SOTA & 22.8\tcite{niu24_chime} & 17.6\tcite{polok2026sedicow} & 20.9\tcite{trelis2026tiron} & \rev{14.5\tcite{tawara2026spokewhenevaluatingspoken}} & -- & \rev{34.0\tcite{tawara2026spokewhenevaluatingspoken}} & -- \\
\bottomrule
\end{tabular}
\vspace{-8pt}
\end{table}

\section{Method}
\label{sec:method}

\subsection{Diarization-conditioned target-speaker ASR}
\label{subsec:recap}
\rev{We build on diarization-conditioned \tsasr{}~\cite{dicow}, using by default the \dicow{}~v3.3 baseline of Self-Enrolled \dicow{} (SE-\dicow{})~\cite{polok2026sedicow}, to which we refer the reader for details.}
\rev{The diarizer provides the probability $d(s,t)\in[0,1]$ that speaker $s\in\{1,\dots,S\}$ is active in frame $t$.}
\rev{For a target speaker $s_k$, these activities define the probabilities of four mutually exclusive frame-level events---silence $\mathcal{S}$ (nobody speaks), target $\mathcal{T}$ (only $s_k$), non-target $\mathcal{N}$ (others, but not $s_k$), and overlap $\mathcal{O}$ ($s_k$ together with others):}
\begin{equation*}
\begin{alignedat}{2}
p_{\mathcal{S}}^t &= \textstyle\prod_{s}(1-d(s,t)), \qquad &
p_{\mathcal{N}}^t &= 1-p_{\mathcal{S}}^t-d(s_k,t),\\
p_{\mathcal{T}}^t &= d(s_k,t)\textstyle\prod_{s\neq s_k}(1-d(s,t)), \qquad &
p_{\mathcal{O}}^t &= d(s_k,t)-p_{\mathcal{T}}^t.
\end{alignedat}
\end{equation*}
\rev{Stacked into $\mathbf{M}^t=\boldsymbol{\phi}_{s_k}(\mathbf{d}_t)=[\,p_{\mathcal{S}}^t\ p_{\mathcal{T}}^t\ p_{\mathcal{N}}^t\ p_{\mathcal{O}}^t\,]^\top$, with $\mathbf{d}_t=[d(s,t)]_{s=1}^{S}$, they form the STNO (silence/target/non-target/overlap) mask, whose size is independent of the number of speakers; the mapping $\boldsymbol{\phi}_{s_k}$ is reused in \cref{sec:csa}.}
This mask modulates the encoder through a frame-level diarization-dependent transformation (FDDT): each frame representation $\mathbf{z}^l_t$ at encoder layer $l$ is mapped by a convex combination of four class-specific affine transforms,
\begin{equation}
\label{eq:FDDT}
\hat{\mathbf{z}}^l_t = \sum_{c\in\{\mathcal{S},\mathcal{T},\mathcal{N},\mathcal{O}\}}
\left(\mathbf{W}_{c}^l \mathbf{z}^l_t + \mathbf{b}_{c}^l\right) p^t_{c},
\end{equation}
\rev{so that a single recognizer, run once per hypothesized speaker, transcribes only speaker $s_k$ while remaining acoustically aware of the others.}
\rev{The transforms $\mathbf{W}_{c}^l$ and $\mathbf{b}_{c}^l$ are trained with the backbone, and in our prior work~\cite{dicow,polok2026sedicow} this lightweight modulation proved more data-efficient and robust than enrollment- or embedding-based conditioning~\cite{ts_asr_kanda,speaker_beam}.}
\rev{By default, each speaker is decoded in a \emph{separate} forward pass, so the passes never exchange information, and the 30\,s windows discard long-range context.}

\subsection{Longform diarization-conditioned Parakeet (\dicop{})}
\label{subsec:longform}
In this work, we integrate the conditioning into the Parakeet FastConformer backbone~\cite{sekoyan2025canary1bv2parakeettdt06bv3efficient,fast_conformer}, which natively decodes long audio in a single pass, and call the result \dicop{} (\cref{fig:system} without the blue boxes).
\rev{The integration keeps the diarization conditioning of \cref{subsec:recap} intact: exactly the same FDDT mechanism (\cref{eq:FDDT}) modulates the encoder, and the only adaptation is subsampling of the STNO mask $\mathbf{M}^t$ to match Parakeet's coarser frame rate.}
Two obstacles arise when the training/decoding chunk grows from 30\,s to several minutes.
First, global self-attention is quadratic in length and exhausts GPU memory.
We use the local (windowed) attention Parakeet supports, bounding memory at the cost of a limited attention span.
\rev{Second, multilingual Parakeet~v3 is trained with the token-and-duration transducer (TDT) loss~\cite{xu_tdt}, a transducer variant whose head predicts, along with each token, how many frames to skip; this head remains the more accurate choice at 30\,s, but computing the transducer \emph{loss} is too expensive for long chunks (\cref{subsec:ctc}).}

\subsection{\rev{From TDT to CTC training}}
\label{subsec:ctc}
\rev{The TDT loss scores the full $T\times U$ lattice of encoder frames and target tokens, so memory grows quadratically with chunk duration.}
\rev{Moreover, the CUDA kernels used to train Parakeet\footnote{NeMo v2.7.3, \path{nemo/collections/asr/parts/numba/rnnt_loss/utils/cuda_utils/gpu_rnnt_kernel.py}} run one thread per target position, so the threads-per-block limit hard-caps the target length at 1024 tokens.}
\rev{Since each thread also holds many registers, the per-block register budget is exhausted well below this cap, and the kernel launch fails with a CUDA error already for shorter targets.}
\rev{A multi-minute chunk exceeds these limits.}
\rev{CTC~\cite{graves_ctc} has neither issue.}
\rev{Parakeet is also released in RNN-T and CTC variants, but these are English-only and less accurate than the TDT models, so we start from the multilingual TDT checkpoint, keep its head \emph{frozen}, and train with a CTC loss instead.}
\rev{The TDT joiner combines an encoder frame $\mathbf{f}_t$ with a predictor state $\mathbf{g}_u$ as}
\begin{equation}
\label{eq:tdt-joint}
\rev{[\mathbf{z}^{\mathrm{tok}}_{t,u};\,\mathbf{z}^{\mathrm{dur}}_{t,u}]
= \mathbf{W}_{o}\,\sigma\!\big(\mathbf{W}_{e}\mathbf{f}_t+\mathbf{W}_{p}\mathbf{g}_u+\mathbf{b}\big),}
\end{equation}
\rev{where $\mathbf{z}^{\mathrm{tok}}_{t,u}$ are the $V{+}1$ token-and-blank logits and $\mathbf{z}^{\mathrm{dur}}_{t,u}$ the duration logits.}
\rev{Zeroing the predictor contribution, $\mathbf{W}_{p}\mathbf{g}_u\equiv\mathbf{0}$, makes the joiner frame-synchronous,}
\begin{equation}
\label{eq:ctc-head}
\rev{P(k\mid t)=\mathrm{softmax}\big(\mathbf{W}^{\mathrm{tok}}_{o}\,\sigma(\mathbf{W}_{e}\mathbf{f}_t+\mathbf{b})\big)_{k},}
\end{equation}
\rev{where $k$ indexes the $V$ tokens and the blank, which is exactly the per-frame posterior CTC needs, while the duration logits are ignored.}
\rev{No parameters are added and TDT decoding remains possible; after CTC finetuning, the posteriors become spiky and blank-dominated, so greedy CTC decoding is much faster and slightly less accurate.}
\rev{The price is a small accuracy degradation relative to TDT training on short chunks, outweighed by the gains from the interaction modules, which need longform training (\cref{sec:results}).}
\rev{All subsequent models build on this CTC variant, whereas the \dicop{} baseline we compare against remains the stronger TDT-trained model.}

\subsection{Cross-speaker attention (\csa{})}
\label{sec:csa}
Lifting the per-speaker independence assumption is the main target of this work.
Since conditioning already runs one pass per speaker, the $S$ streams of a conversation form a natural batch that we let interact inside the encoder (blue boxes in \cref{fig:system}).
\rev{\dicow{}~\cite{dicow} did so through a multichannel-style \emph{co-attention}~\cite{horiguchi22_coattention} over time and channels, which in later shortform experiments did not improve over DiCoW~v3.3 alone while incurring a substantial GPU-memory overhead.}
\rev{Since our streams share one encoder and one time axis, we attend only across the speaker axis at each frame instead.}

\textbf{Cross-speaker attention.}
Let $\mathbf{z}^{l}_{s,t}\in\mathbb{R}^{D}$ denote the FDDT-modulated state (\cref{eq:FDDT}) of speaker $s$ at layer $l$ and frame $t$, and let $\mathbf{Z}^{l}_{t}=[\mathbf{z}^{l}_{1,t},\dots,\mathbf{z}^{l}_{S,t}]^{\top}\in\mathbb{R}^{S\times D}$ stack the conversation's $S$ speakers at that frame.
At selected layers we run multi-head attention along the \emph{speaker} axis,
\begin{equation}
\label{eq:csa-attn}
\mathbf{U}^{l}_{t}=\mathrm{MHA}\!\big(\mathrm{LN}(\mathbf{Z}^{l}_{t})\big),
\end{equation}
where query, key, and value are all $\mathrm{LN}(\mathbf{Z}^{l}_{t})$, and add the result back \rev{through a plain residual},
\begin{equation}
\label{eq:csa-res}
\rev{\mathbf{z}^{l}_{s,t}\;\leftarrow\;\mathbf{z}^{l}_{s,t}+\mathbf{u}^{l}_{s,t}.}
\end{equation}
\rev{The output projection of the attention is initialized close to zero (std $10^{-4}$), so \csa{} is initially a near-identity and a warm-started \dicop{} encoder is left almost unchanged.}
\rev{As $S$ acts as the sequence length without positional encoding, the number of speakers is not fixed.}

\textbf{STNO mask refiner.}
A limitation of the STNO conditioning is that speakers overlapping in a frame receive the \emph{same} mask---an overlap frame is labeled identically for all of them---so the conditioning alone cannot tell them apart.
Longform decoding should already help here: similarly to SE-DiCoW~\cite{polok2026sedicow}, per-speaker evidence can be aggregated across the whole recording into the global tokens.
We additionally let the model re-estimate a soft per-speaker activity from the cross-speaker-informed states,
\begin{equation}
\label{eq:refine-act}
\hat{d}^{l}(s,t)=\sigma\!\big(\rev{\mathbf{w}_{l}^{\top}}\mathrm{LN}(\mathbf{z}^{l}_{s,t})+\rev{b_{l}}\big)\in[0,1],
\end{equation}
\rev{stacks the $S$ re-estimated activities into $\hat{\mathbf{d}}^{l}_{t}=[\hat{d}^{l}(s,t)]_{s=1}^{S}$, rebuilds a fresh STNO mask $\hat{\mathbf{M}}^{l,t}=\boldsymbol{\phi}_{s_k}(\hat{\mathbf{d}}^{l}_{t})$ with the mapping of \cref{subsec:recap}, and blends it back through a sigmoid gate,}
\begin{equation}
\label{eq:refine-blend}
\mathbf{M}^{t}\;\leftarrow\;(1-\beta_{l})\,\mathbf{M}^{t}+\beta_{l}\,\hat{\mathbf{M}}^{l,t},
\qquad \beta_{l}=\sigma(\rev{g_{l}}).
\end{equation}
\rev{Each refiner has its own parameters $\mathbf{w}_{l}$, $b_{l}$, and $g_{l}$.}
\rev{The logits $g_{l}$ start negative, so the input mask is unchanged at initialization.}

\begin{figure}[t]
\centering
\pgfplotsset{modelmarks/.style={
  scatter, scatter src=explicit symbolic,
  scatter/classes={
    se={mark=diamond*,  mark size=2.2pt},
    dw={mark=*,         mark size=1.8pt},
    gl={mark=square*,   mark size=1.8pt},
    lt={mark=triangle*, mark size=2.2pt},
    pd={mark=star,      mark size=3.2pt, line width=0.9pt},
    lc={mark=pentagon*, mark size=2.2pt}}}}
\begin{tikzpicture}
\begin{axis}[
  width=\columnwidth, height=5.0cm,
  xmode=log, log ticks with fixed point,
  xmin=30, xmax=620, ymin=10.6, ymax=19.6,
  xtick={30,50,100,200,300,500},
  xminorticks=false,
  xlabel={RTFx (log scale)\quad faster $\rightarrow$},
  ylabel={\cpwer{} [\%]},
  xlabel style={font=\small, yshift=2pt}, ylabel style={font=\small},
  ytick={11,13,...,19},
  tick label style={font=\scriptsize},
  grid=major, grid style={draw=black!12, line width=0.3pt},
  axis line style={line width=0.4pt},
  legend style={at={(0.5,1.02)}, anchor=south, font=\scriptsize, draw=none, fill=none,
                legend columns=3, /tikz/every even column/.append style={column sep=4pt}},
  legend cell align=left,
  every axis plot/.append style={line width=0.7pt},
  clip=false,
]
\addlegendimage{only marks, mark=diamond*,  mark size=2.2pt, black} \addlegendentry{SE-\dicow{}}
\addlegendimage{only marks, mark=*,         mark size=1.8pt, black} \addlegendentry{\dicow{} v3.3}
\addlegendimage{only marks, mark=square*,   mark size=1.8pt, black} \addlegendentry{\dicop{} glob/TDT}
\addlegendimage{only marks, mark=triangle*, mark size=2.2pt, black} \addlegendentry{\dicop{} loc/TDT}
\addlegendimage{only marks, mark=pentagon*, mark size=2.2pt, black} \addlegendentry{\dicop{} loc/CTC}
\addlegendimage{only marks, mark=star, mark size=3.2pt, line width=0.9pt, black} \addlegendentry{Pandemonium}
\addplot[cNSF, modelmarks] coordinates {(37,15.3) [se] (54,15.5) [dw] (201,18.7) [gl] (269,17.3) [lt] (330,14.1) [pd] (405,18.8) [lc]};
\addplot[cAMI, modelmarks] coordinates {(37,14.4) [se] (54,14.5) [dw] (201,16.0) [gl] (269,14.5) [lt] (330,13.4) [pd] (405,15.1) [lc]};
\addplot[cMx6, modelmarks] coordinates {(37,11.7) [se] (54,11.4) [dw] (201,12.7) [gl] (269,11.9) [lt] (330,11.1) [pd] (405,12.6) [lc]};
\addplot[cSBC, modelmarks] coordinates {(37,15.2) [se] (54,15.0) [dw] (201,18.8) [gl] (269,16.0) [lt] (330,13.8) [pd] (405,17.0) [lc]};
\node[anchor=west, font=\scriptsize, cNSF] at (axis cs:450,18.8) {NSF};
\node[anchor=west, font=\scriptsize, cSBC] at (axis cs:450,17.0) {SBC};
\node[anchor=west, font=\scriptsize, cAMI] at (axis cs:450,15.1) {AMI};
\node[anchor=west, font=\scriptsize, cMx6] at (axis cs:450,12.6) {Mx6};
\end{axis}
\end{tikzpicture}
\vspace{-9pt}
\caption{\rev{Accuracy/speed trade-off per evaluation set (oracle diarization) against real-time-factor speedup (RTFx, greedy decoding).
Line color: evaluation set; marker: model, connected in order of speed; all decode full conversations.
All systems but \emph{Pandemonium} train on the \emph{DiCoW-orig.} data (\cref{subsec:datasets}); \dicop{} variants differ in attention (\emph{loc}/\emph{glob}: local/global) and training loss, without \csa{}.}}
\label{fig:tradeoff}
\end{figure}
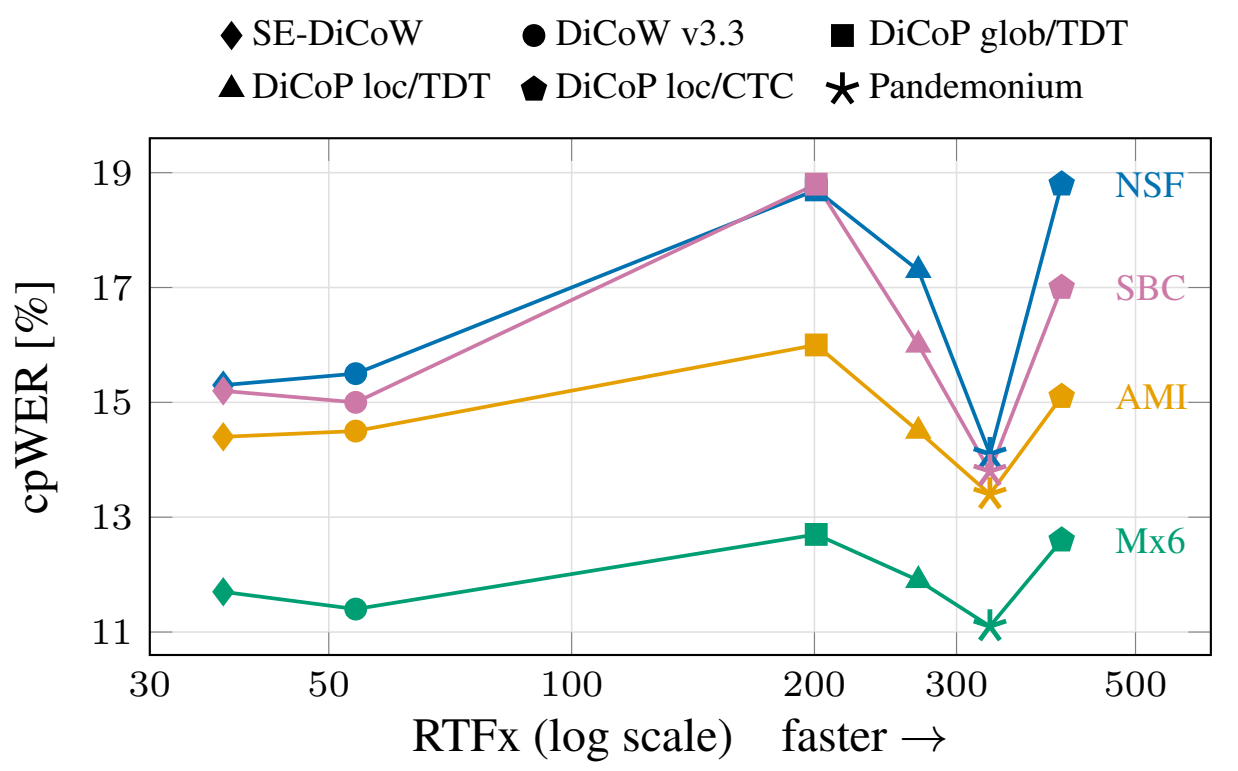

\section{Experimental Setup}
\label{sec:setup}
\rev{Throughout this paper, we report the concatenated minimum-permutation word error rate (\cpwer{})~\cite{watanabe2020chime}, computed with MeetEval~\cite{MeetEval23}, with oracle diarization unless stated otherwise.}
\rev{Predicted diarization comes from pyannoteAI \mbox{Precision-3}}~\cite{laurent26_odyssey,bredin23_interspeech}.\footnote{\url{https://www.pyannote.ai/blog/precision-3}}
For consistency, all models use greedy decoding.

\subsection{Datasets}
\label{subsec:datasets}
We use four sources of training data.
\rev{\emph{Real} data is the genuine-meeting subset (115\,h): the single-distant-microphone training sets of NOTSOFAR-1~\cite{abramovski2025summarynotsofar1challengehighlights} (35.6\,h) and AMI~\cite{Mccowan2005_ami} (79.4\,h) only.}
\rev{\emph{DiCoW-orig.} data (1.2k\,h) is the original SE-DiCoW training recipe~\cite{polok2026sedicow}---AMI, NOTSOFAR-1, LibriMix~\cite{Cosentino2020LibriMixAO}, and LibriSpeechMix-like mixtures---and drives the controlled ablation (\cref{fig:tradeoff}).}
\emph{Synthetic} data (4k\,h in total) serve as initialization: 2k\,h of FastMSS~\cite{synthetic} mixtures of up to eight speakers from LibriSpeech~\cite{librispeech}, and 2k\,h built from Fisher~\cite{fisher} with Sidon~\cite{sidon} and forced alignment~\cite{mcauliffe17_interspeech}.
\rev{\emph{Pseudo-labels} are the filtered YODAS2 self-labels described below (5.4k\,h mined, of which 4.6k\,h train the Parakeet-based models).}
We evaluate on NOTSOFAR-1 (NSF), AMI, ICSI~\cite{ICSI}, AISHELL-4 (A4)~\cite{fu21b_aishell}, AliMeeting (Ali)~\cite{Yu2021M2MetTI}, the Santa Barbara Corpus (SBC)~\cite{sbc}, Mixer6 (Mx6)~\cite{Mixer6}, and DiPCo~\cite{segbroeck20_dipco}.
\rev{For SBC we use the conversation and task subsets, restricted to recordings with two to eight speakers.}
Tables and figures use these abbreviations.
\subsubsection{Pseudo-labeling YODAS2}
\label{sec:pseudo}
\rev{Since genuine multi-speaker interaction is scarce in curated data and unrealistic in synthetic data, we pseudo-label the YODAS2 corpus~\cite{yodas}.}
We diarize it with pyannoteAI \mbox{Precision-2}~\cite{laurent26_odyssey, bredin23_interspeech} and, among the 10\,min chunks that lie within the Whisper languages and contain at most eight speakers, select \rev{5.4k\,h with the highest speaker overlap}.
\rev{Final pseudo-labels are produced by re-diarizing the selected data with an in-house eight-speaker streaming Sortformer~\cite{medennikov25_interspeech} and transcribing with DiCoW~v3.3~\cite{polok2026sedicow} conditioned on that diarization.}
\rev{We then clean the labels with three filters: a language-ID check with Kimi~K3~\cite{kimi_k3} against the conditioning language, a regex detector of $n$-gram repetitions (hallucinations), and a leakage filter that discards segments where two speakers share five or more consecutive words simultaneously.}
\rev{Of the 5.4k\,h mined, \dicop{} and \emph{Pandemonium} train on the 4.6k\,h that remain after dropping 0.8k\,h in languages Parakeet does not support (e.g., Mandarin).}
The \dicow{} pseudo-label study of \cref{tab:yodas-filter-full} instead uses the 3.4k\,h left after segmenting into 30\,s windows and dropping single-speaker ones.

\begin{table}[t]
\centering
\caption{\rev{Influence of pseudo-label quality on meeting benchmarks.}
\rev{Rows: \dicow{} with each cleaning filter applied to the full pool; orig.: trained on the \emph{DiCoW-orig.} data.}
\cpwer{}~[\%]~$\downarrow$.
Best per column in \textbf{bold}.
\underline{Underlined}: in-domain, i.e., the matched training set was used.}
\label{tab:yodas-filter-full}
\setlength{\tabcolsep}{2.5pt}
\footnotesize
  \begin{tabular}{@{}l r cccccc c@{}}
  \toprule
  System / selection & Hrs & NSF & AMI & A4 & Ali & Mx6 & SBC & Avg \\
  \midrule
  \dicow{} v3.3 (orig.)~\cite{polok2026sedicow}  & 1.2k & \underline{15.5} & \underline{14.5} & 33.7 & 39.3 & 11.4 & \textbf{15.0} & 21.6 \\
  SE-\dicow{} (orig.)~\cite{polok2026sedicow}    & 1.2k & \underline{\textbf{15.3}} & \underline{\textbf{14.4}} & 40.2 & 43.4 & 11.7 & 15.2 & 23.4 \\
  \dicow{} (synthetic)~\cite{synthetic}   & 4.0k & 23.8 & 24.7 & 26.8 & 35.9 & 16.8 & 22.1 & 25.0 \\
  \midrule
  \multicolumn{9}{@{}l}{\emph{\dicow{} --- pseudo-labels only (no real data):}} \\
  All pseudo-labels          & 3.4k & 17.2 & 19.1 & 24.3 & 31.3 & 11.7 & 15.7 & 19.9 \\
  \;\; LID check (only)      & 3.1k & 17.3 & 19.4 & 24.5 & 31.4 & 11.8 & 16.3 & 20.1 \\
  \;\; hallucination (only)  & 3.2k & 17.1 & 18.0 & 24.4 & 31.4 & 11.2 & 15.5 & 19.6 \\
  \;\; leakage filter (only) & 3.3k & 17.0 & 18.9 & 24.6 & 31.6 & 14.2 & 16.0 & 20.4 \\
  \;\; all filters combined  & 2.9k & 17.0 & 18.3 & 24.0 & 30.9 & \textbf{10.9} & 15.7 & 19.5 \\
  \midrule
  \multicolumn{9}{@{}l}{\emph{\dicow{} --- pseudo-labels + finetune on real:}} \\
  \;\;FT: DiCoW-orig.        & 4.1k & \underline{15.5} & \underline{\textbf{14.4}} & 28.1 & 37.1 & 11.4 & 15.1 & 20.3 \\
  \;\;FT: NSF+AMI+A4+Ali     & 3.2k & \underline{15.8} & \underline{15.0} & \underline{\textbf{15.6}} & \underline{\textbf{19.3}} & 11.7 & 15.3 & \textbf{15.5} \\
  \bottomrule
  \end{tabular}
\vspace{-8pt}
\end{table}

\subsection{Model and training details}
\label{subsec:model-and-training-details}
\dicop{} warm-starts from Parakeet-TDT-0.6b-v3~\cite{sekoyan2025canary1bv2parakeettdt06bv3efficient} with local attention (context $[256,256]$), decoded by the TDT head or the CTC head of \cref{subsec:ctc}.
\emph{Pandemonium} inserts cross-speaker attention after encoder layers $\{8,12,16,20\}$ and the STNO mask refiner after $\{12,16,20\}$\rev{; training conversations are capped at $S=8$ speakers}.
\rev{The diarization mask is randomly corrupted in training so the refiner has errors to correct and gets a $10\times$ learning rate.}
\rev{Configurations are released with the models; \dicow{} baselines follow SE-DiCoW~\cite{polok2026sedicow}.}

\begin{table}[t]
\centering
\caption{Ablation of \emph{Pandemonium}.
\emph{Top:} training data.
\emph{Bottom:} architecture, \rev{removing the STNO refiner, then also \csa{} (plain \dicop{}), and the full model trained on 30\,s windows}.
\cpwer{}~[\%]~$\downarrow$.
}
\label{tab:ablation}
\setlength{\tabcolsep}{4pt}
\footnotesize
\begin{tabular}{@{}l cc cc c@{}}
\toprule
 & \multicolumn{2}{c}{In-domain} & \multicolumn{2}{c}{Out-of-domain} & \\
\cmidrule(lr){2-3}\cmidrule(lr){4-5}
System & NSF & AMI & Mx6 & SBC & Avg \\
\midrule
\textbf{Pandemonium} (full)              & 14.1 & 13.4 & 11.1 & 13.8 & 13.1 \\
\midrule
\multicolumn{6}{@{}l}{\emph{Training-data source:}} \\
\;\; w/o real (synth+pseudo)             & 14.4 & 18.1 & 17.7 & 14.0 & 16.1 \\
\;\; pseudo-labels only                  & 20.2 & 26.6 & 16.8 & 16.9 & 20.1 \\
\;\; synthetic only                      & 16.7 & 17.6 & 10.9 & 15.0 & 15.0 \\
\midrule
\multicolumn{6}{@{}l}{\emph{Architecture} (synthetic-only pretraining)\emph{:}} \\
\;\; $-$\,refiner                        & 18.8 & 19.6 & 11.7 & 17.0 & 16.8 \\
\;\; $-$\,refiner $-$\,\csa{} (\dicop{}) & 19.5 & 20.0 & 11.8 & 17.1 & 17.1 \\
\;\; full, 30\,s windows                 & 19.3 & 20.0 & 11.9 & 17.1 & 17.1 \\
\bottomrule
\end{tabular}
\vspace{-10pt}
\end{table}

\section{Results and Discussion}
\label{sec:results}

\Cref{tab:main} presents our main results.
\rev{\emph{Pandemonium} outperforms the DiCoW~v3.3 baseline~\cite{polok2026sedicow} and the longform \dicop{} backbone on every benchmark with oracle diarization, and on all but two with predicted diarization.}
\rev{With predicted diarization, it also surpasses the published single-microphone state-of-the-art.}
\rev{Without multichannel processing, decoding every distant microphone and merging the hypotheses with MOVER~\cite{kamo25_interspeech} (\emph{Pandemonium}\,+\,MOVER) lowers \cpwer{} by a further 0.8--3.4 absolute.}
\rev{Note that the \dicop{} row is the stronger TDT-trained model, favoring the baseline.}

\rev{\textbf{A fast but weaker baseline.}}
\rev{\Cref{fig:tradeoff} establishes the longform \tsasr{} baseline: integrating the conditioning into Parakeet yields a model that decodes 4--8$\times$ faster than \dicow{} but is up to 3.8 \cpwer{} worse.}
\rev{Local attention decodes full conversations faster and better than global attention on every set, so we adopt it for longform.}
\rev{Switching to CTC training costs 0.6--1.5 \cpwer{} but raises the RTFx by $1.5\times$ (269 to 405).}
\rev{Longer training chunks (60\,s, 10\,min; not shown) do not close the gap, moving \cpwer{} by at most 0.6 in either direction.}
\rev{\emph{Pandemonium} keeps most of this speed (RTFx 330 vs.\ 405) and is the most accurate on every set.}

\rev{\textbf{Pseudo-labels as a source of interaction.}}
\rev{Curated meeting corpora contain genuine multi-speaker interaction but are far too small to learn it from, so we mine more from YODAS2 (\cref{sec:pseudo}) and first check the quality of the resulting pseudo-labels by training \dicow{} on them alone (\cref{tab:yodas-filter-full}).}
\rev{Even unfiltered, a \dicow{} trained on this data is competitive with both the synthetic and the \dicow{}~v3.3 (orig.) reference, and reaches a better average \cpwer{} (19.9 vs.\ 21.6).}
Thanks to the diverse YODAS2 languages, the pseudo-label-trained model is more generic than the English-skewed v3.3.
\rev{SE-\dicow{}, in turn, matches v3.3 in-domain but is 4--7 \cpwer{} worse on the Mandarin sets, so self-enrollment does not transfer across languages.}
Among the individual filters, only hallucination removal gives a clear improvement.
We attribute the robustness to the remaining label noise to the diversity of the data, which, together with batch size $96$ and gradient clipping, averages the noise out and skips the worst examples, as also observed in OWLS~\cite{chen2025owls}.
\rev{Finally, finetuning the pseudo-label-pretrained checkpoint on the \emph{DiCoW-orig.} data improves NSF and AMI but degrades AISHELL-4 and AliMeeting.}
\rev{Finetuning on the real meeting sets alone, with the Mandarin AISHELL-4 and AliMeeting training sets added, instead brings both to 15.6 and 19.3 \cpwer{}, far below any English-only model, and gives the best average, 15.5.}

\rev{\textbf{Interactions turn the baseline around.}}
\Cref{tab:ablation} isolates each contribution of \emph{Pandemonium}.
\rev{The interaction modules of \cref{sec:csa} help: over plain \dicop{} ($17.1$ \cpwer{}), \csa{} (\cref{eq:csa-attn}) gains $0.3$ and the STNO mask refiner (\cref{eq:refine-blend}) a further $1.8$ ($15.0$).}
\rev{They need longform context, though: the same architecture trained on 30\,s windows falls back to $17.1$, forfeiting the entire gain.}
\rev{On the \emph{data} side, the picture is clear: the three-stage schedule synthetic\,$\to$\,pseudo\,$\to$\,real is best, with synthetic mixtures serving as a useful initialization and real data as a final refinement.}
Pseudo-labels alone trail synthetic-only ($20.1$ vs.\ $15.0$).
We attribute this to the English-skewed evaluation, which hides the cross-lingual gains of the pseudo-labels, as Parakeet does not support Mandarin.
\rev{Together, they take \dicop{} from 16.7 to 14.7 average \cpwer{} in \cref{tab:main}.}

\section{Conclusion}
\label{sec:conclusion}
We addressed two limitations of diarization-conditioned \tsasr{}: independent per-speaker decoding and the fixed 30\,s window.
\rev{Integrating the conditioning into a longform Parakeet backbone gives a fast but weaker baseline; cross-speaker attention and an STNO mask refiner, trained on 4.6k hours of pseudo-labeled \mbox{YODAS2} conversations, make it state-of-the-art.}
\rev{Future work will fuse \dicop{} with a conversation-pretrained language model, recover TDT training via a pruned transducer loss, and train diarization and recognition jointly.}

\bibliographystyle{IEEEbib}
\bibliography{references}

\end{document}